\documentclass[twocolumn,twocolappendix]{aastex63}
\usepackage{amsmath}

\usepackage{lineno}

\newcommand{\be}{\begin{eqnarray}}
\newcommand{\ee}{\end{eqnarray}}

\renewcommand{\vec}[1]{\boldsymbol{#1}}

\newcommand\js{\bgroup\markoverwith{\textcolor[rgb]{0.1, .5, .1}{\rule[0.5ex]{8pt}{1.5pt}}}\ULon}
\newcommand\as{\bgroup\markoverwith{\textcolor[rgb]{.5, 0, .6}{\rule[0.5ex]{8pt}{1.5pt}}}\ULon}

\submitjournal{ApJL} 

\shorttitle{BH Binary Formation}
\shortauthors{Li et al.}

\begin{document}

\title{Hydrodynamical Simulations of Black-Hole Binary Formation in AGN Disks} 

\correspondingauthor{Jiaru Li}
\email{jiaru\textunderscore li@astro.cornell.edu}

\author[0000-0001-5550-7421]{Jiaru Li}
\affiliation{Theoretical Division, Los Alamos National Laboratory, Los Alamos, NM 87545, USA}
\affiliation{Center for Astrophysics and Planetary Science,
Department of Astronomy, Cornell University, Ithaca, NY 14853, USA}

\author[0000-0001-8291-2625]{Adam M. Dempsey}
\affiliation{X-Computational Physics Division, Los Alamos National Laboratory, Los Alamos, NM 87545, USA}

\author[0000-0003-3556-6568]{Hui Li}
\affiliation{Theoretical Division, Los Alamos National Laboratory, Los Alamos, NM 87545, USA}

\author[0000-0002-1934-6250]{Dong Lai}
\affiliation{Center for Astrophysics and Planetary Science, Department of Astronomy, Cornell University, Ithaca, NY 14853, USA}

\author[0000-0002-4142-3080]{Shengtai Li}
\affiliation{Theoretical Division, Los Alamos National Laboratory, Los Alamos, NM 87545, USA}

\begin{abstract}
We study the close encounters between two single black holes (BHs) embedded in an AGN disk using a series of global 2D hydrodynamics simulations.
We find that when the disk density is sufficiently high, bound BH binaries can be formed by the collision of their circum-single disks. 
Our analysis demonstrates that, after a BH pair passes the pericenter of their relative trajectory, a gas post-collision drag may slow down the BHs, possibly forcing the two BHs to stay tightly bound.
A binary formed by a close encounter can have a compact semi-major axis, large eccentricity, and retrograde orbital angular momentum.
We provide a fitting formula that can accurately predict whether a close encounter can form a binary based on the gas mass and the incoming energy of the encounter.
This fitting formula can be easily implemented in other long-term simulations that study the dynamical evolution of BHs in AGN disks.
\end{abstract}


\section{Introduction}  

\label{sec:intro}
Mergers of black holes (BH) are the most common gravitational wave (GW) sources for LIGO/Virgo to date \citep{LIGO2019, LIGO2021a, LIGO2021b}.
The large masses of some detected merger events suggest that some sources may be the result of repeated BH mergers.
While globular clusters are a well-studied breeding ground for these heavyweight events, another intriguing venue is the accretion disks around active galactic nuclei (AGN). 
\citep[e.g.,][]{McKernan2012, McKernan2014, Bartos2017, Stone2017, Secunda2019, Secunda2020, Yang2019a, Yang2019b, Grobner2020, Ishibashi2020, Tagawa2020b, Tagawa2020a, Ford2021, Boekholt2022}.
Mergers of binary BHs (BBHs) embedded in AGN disks have been argued to have distinct observable distributions of mass, spin, and eccentricity \citep[e.g.,][]{McKernan2018, Yang2019a,Gerosa2021,LiG2022,LJR2022ApJ}.
Because of the gas-rich environment, these mergers may also produce electromagnetic counterparts that can be associated with their GW signals \citep{Stone2017, McKernan2019, Graham2020,Palmese2021Apj}.

In recent years, several mechanisms for BBH hardening in AGN disks have been proposed.
For example, binaries may be hardened or even driven to eccentric mergers through a series of nearly co-planar binary-single scatterings \citep[e.g.,][]{Leigh2018, Samsing2020}.
Under some special conditions, binaries can also acquire large eccentricities (and hence small pericenter separations) if they are captured into evection resonances as they migrate radially within the AGN disk \citep{Bhaskar2022ApJ,Munoz2022arxiv}.
As binaries perturb the gas distribution around them through gravity, the gas can also affect the binary orbit \citep[e.g.,][]{Baruteau2011,Stone2017}. 
Recent high-resolution hydrodynamics simulations have carefully diagnosed the binary-disk interaction and have shown that the interaction can lead to the contraction of embedded binaries for a wide range of binary parameters,  varieties of thermal properties of the AGN disks, and different accretion prescriptions \citep[][]{LiYP2021ApJ,LiYP2022ApJL,Dempsey2022arxiv,LiRX2022arxiv02,LiRX2022arxiv07}.

While all mechanisms above assume pre-existing BBHs, an open question 
remains as to how these binaries actually form.
Although some binaries could be formed in nuclear star clusters and captured into the AGN disk \citep{Bartos2017}, it is often assumed that many binaries are dynamically assembled through close encounters between BHs that are born in the disk \citep[e.g.,][]{Stone2017}.
As AGN disks may have non-monotonic radial profiles \citep{Sirko2003, Thompson2005, Dittmann2020MNRAS}, embedded BHs may naturally approach each other due to convergent migration or migration traps \citep{Bellovary2016}.
Tightly-packed BH orbits will then lead to more frequent close encounters and opportunities for binary formation.

The literature that analyzes the binary formation process during a close encounter is still limited.
\cite{Secunda2019,Secunda2020} showed with $N$-body simulations that migration traps can enhance the rate of close encounters. 
However, the vast majority of these close encounters only lead to weakly bound and short-lived binaries \citep{LJR2022ApJ}. 
Although these temporary binaries might be further hardened by three-body scatterings \citep{Heggie1975MNRAS,Stone2017}, the allowed time window for the follow-up hardening is very short because the probability for two encountering BHs to hold as a (temporary) binary diminishes exponentially with the binary lifetime \citep{Boekholt2022}. 
\cite{Tagawa2020a} suggested that dynamical friction from the AGN disk gas can aid in binary capture, but their results were based on one-dimensional simulations and assumed circular orbits for all BBHs. 

By performing a large number of $N$-body simulations of BH close encounters in AGN disks, \cite{LJR2022ApJ} found that, even without any gas dynamical effects (e.g. when the AGN disk density is sufficiently small), the relative distance between two BHs can become arbitrarily small during a close encounter, enhancing the probability of capture via GW emission.
\cite{Boekholt2022} also found similar results using a phase space fractal structure analysis. 
However, that paper did not implement the gas force onto the BHs from the AGN disk. \cite{LJR2022ApJ} adopted a simple drag force model 
to mimic the gas effect; they found for a drag time of $10^5$ orbital periods or more (corresponding to relatively low/modest AGN disk density), the drag force
does not lead to an enhanced binary formation rate. However, it is not clear that linear gas drag \citep{Papaloizou2000MNRAS} captures the full hydrodynamical effects of AGN disks. 
In fact, we will show that drag mechanism is highly non-linear.

In this Letter, we study the close encounters between two single BHs embedded in an AGN disk using a series of global 2D hydrodynamics simulations (presented in Sections \ref{sec:method}-\ref{sec:result}).
We show that collisions of the disks around each BH help dissipate BH kinetic energy via a drag-like effect.
We find that bound binaries can form when the disk mass is sufficiently high. Our mechanism does not require GW emission or three-body scattering.
Finally, we provide a simple model (fitting formula) that can be implemented in $N$-body simulations that determine the outcomes of close encounters
(Section \ref{sec:conditions}).

\section{Method}
\label{sec:method}

\begin{figure*}[t]
    \epsscale{1.1}
    \plotone{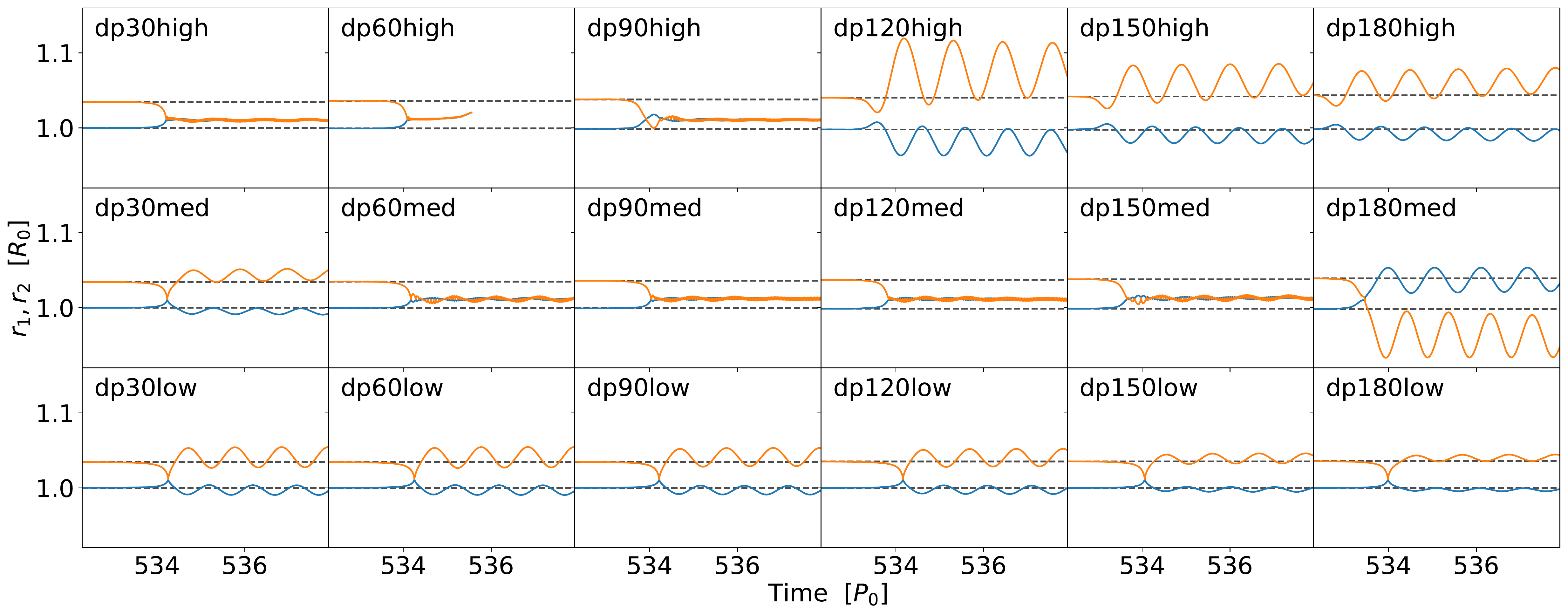}
    \caption{Evolution of $r_1$ (blue) and $r_2$ (orange) around the time when the BHs encounter each other for our main suite of simulations. Simulations are grouped by disk mass (rows; increasing from bottom to top, with $\Sigma_0R_0^2/M=10^{-5}$, $5\times10^{-5}$ and $10^{-4}$) and initial azimuthal separation of the two BHs $\Delta\varphi$ (columns; increasing from left to right, with $\Delta\varphi=30^\circ$, $60^\circ$, $90^\circ$, $120^\circ$, $150^\circ$, and $180^{\circ}$). Generally, we find that the likelihood of binary capture increases as the disk mass increases. But large initial azimuthal separation can hinder formation. 
    }
    \label{fig:r12_table}
\end{figure*}
\begin{figure*}[htb]
    \epsscale{1.0}
    \plotone{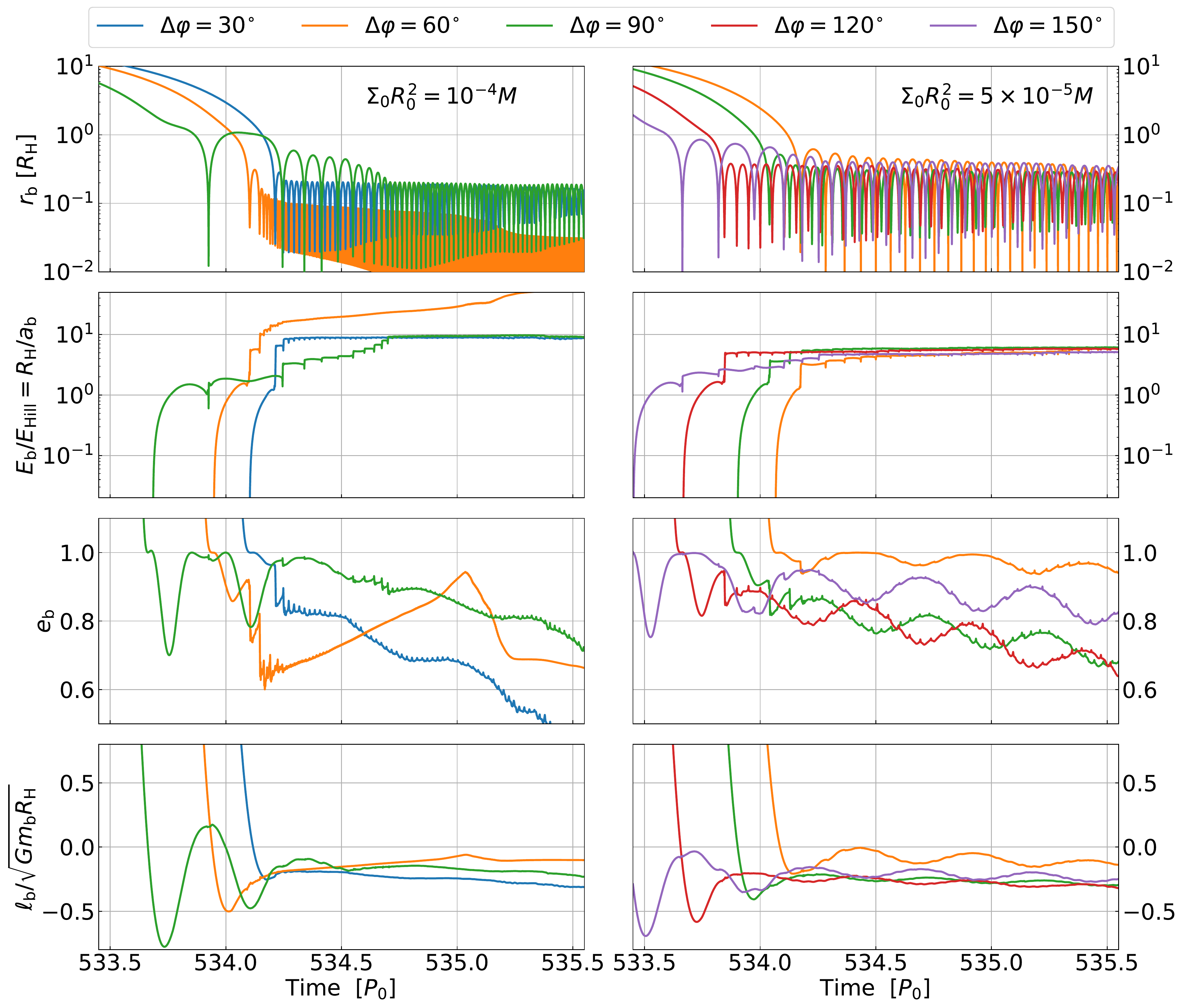}
    \caption{
    Orbital evolution of all bound binaries in our main suite of simulations. From top to bottom we plot the time evolution of BH separation, binding energy in units of $E_{\rm Hill}=-Gm_b/(2R_{\rm H})$ (or equivalently the ratio of $R_{\rm H}$ to the binary semi-major axis $a_{\rm b}$), eccentricity, and specific angular momentum. Binaries form via multiple close encounters that collectively dissipate a large amount of the BHs' kinetic energy. Once formed, all binaries are retrograde and eccentric. Over time, their eccentricity damps and they tighten to our resolution limit. 
    }
    \label{fig:rael-vs-time}
\end{figure*}
\begin{figure*}[t]
    \epsscale{0.4116}
    \plotone{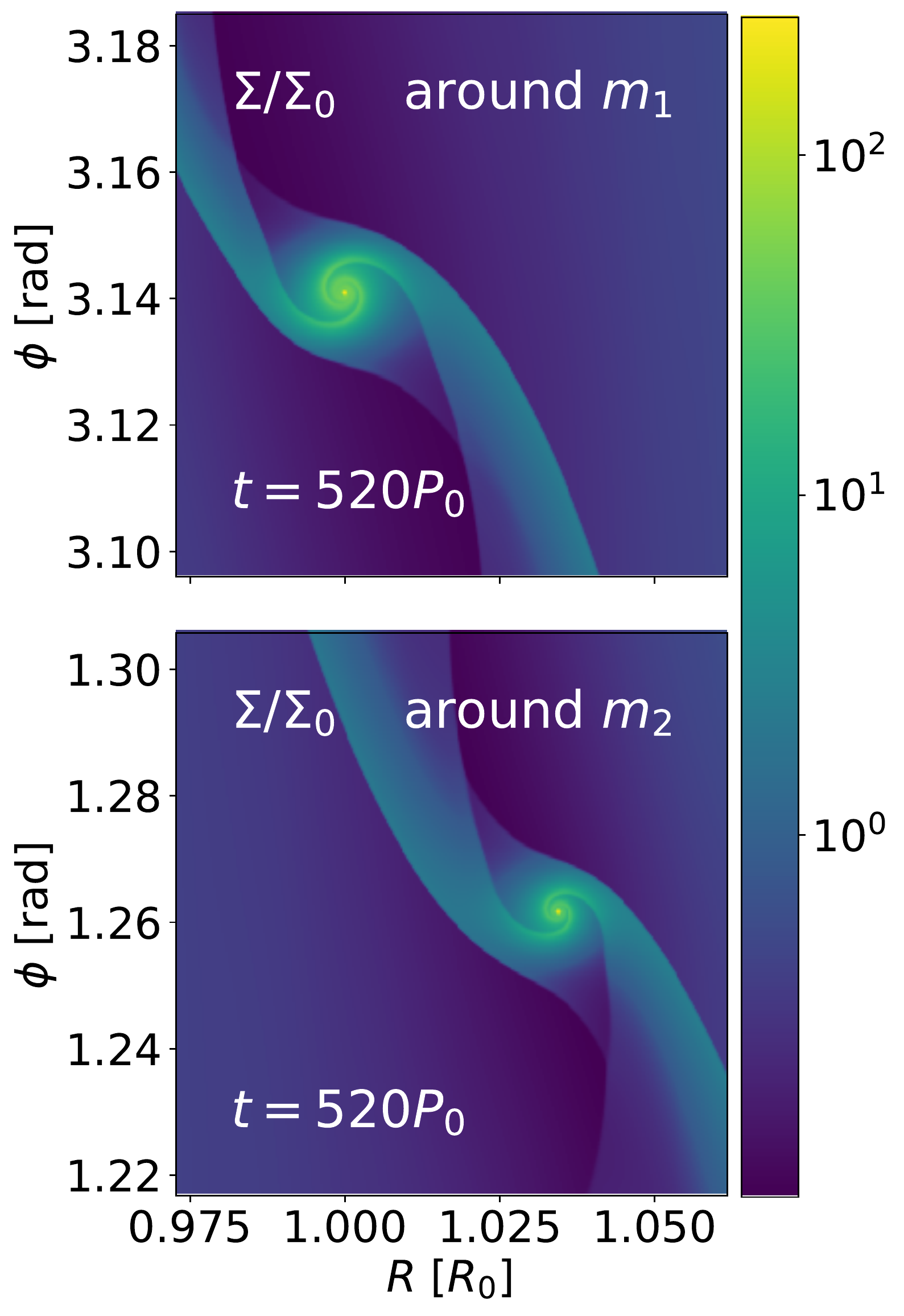}
    \epsscale{0.732}
    \plotone{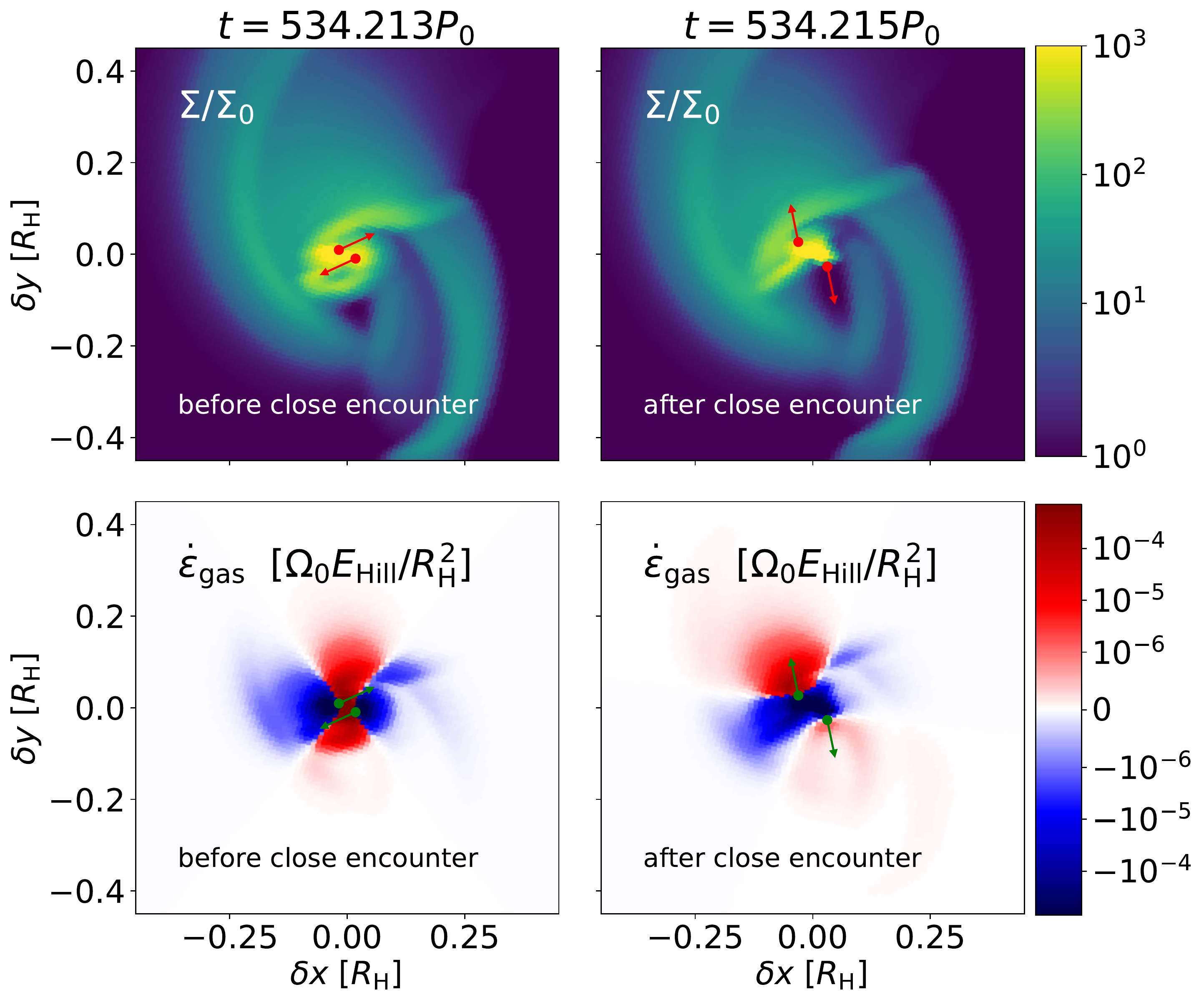}
    \caption{
    Snapshots of the gas surface density and power per unit area delivered to the BHs for the exemplary simulation \texttt{dp30high} at three different times before and after the first close encounter. At early times (left column), each BH supports a circum-single disk (CSD) and trailing Lindblad spirals. Just before close encounter, the CSDs collide (middle column). As the BHs depart (right column), the CSD gas is left behind and drags the BHs ($\dot{\varepsilon}_{\rm gas}<0$), slowing them down. }
    \label{fig:2d-den-Edot}
\end{figure*}
\begin{figure*}[t]
    \centering
    \epsscale{0.53}
    \plotone{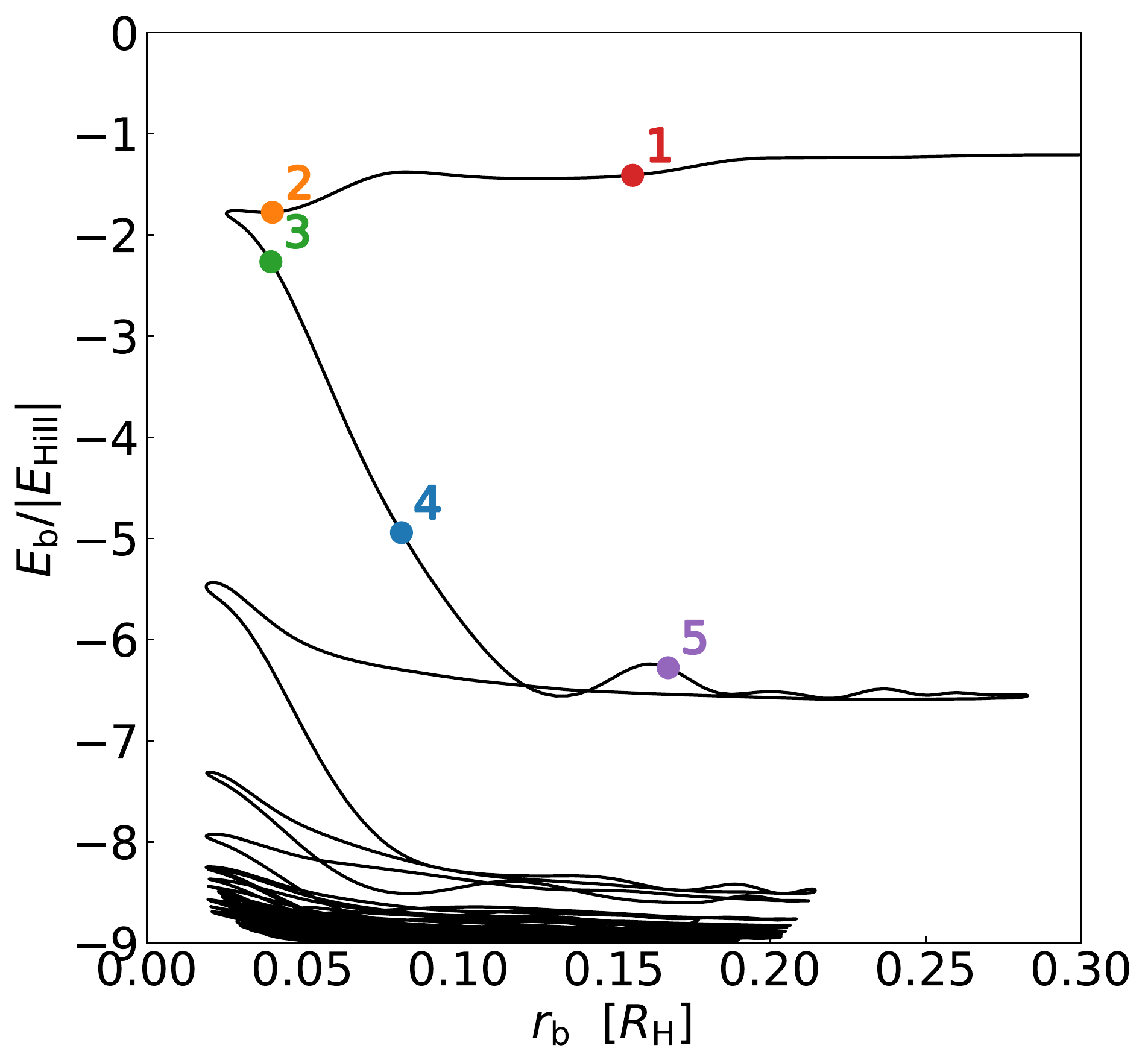}\\
    \epsscale{0.5}
    \plotone{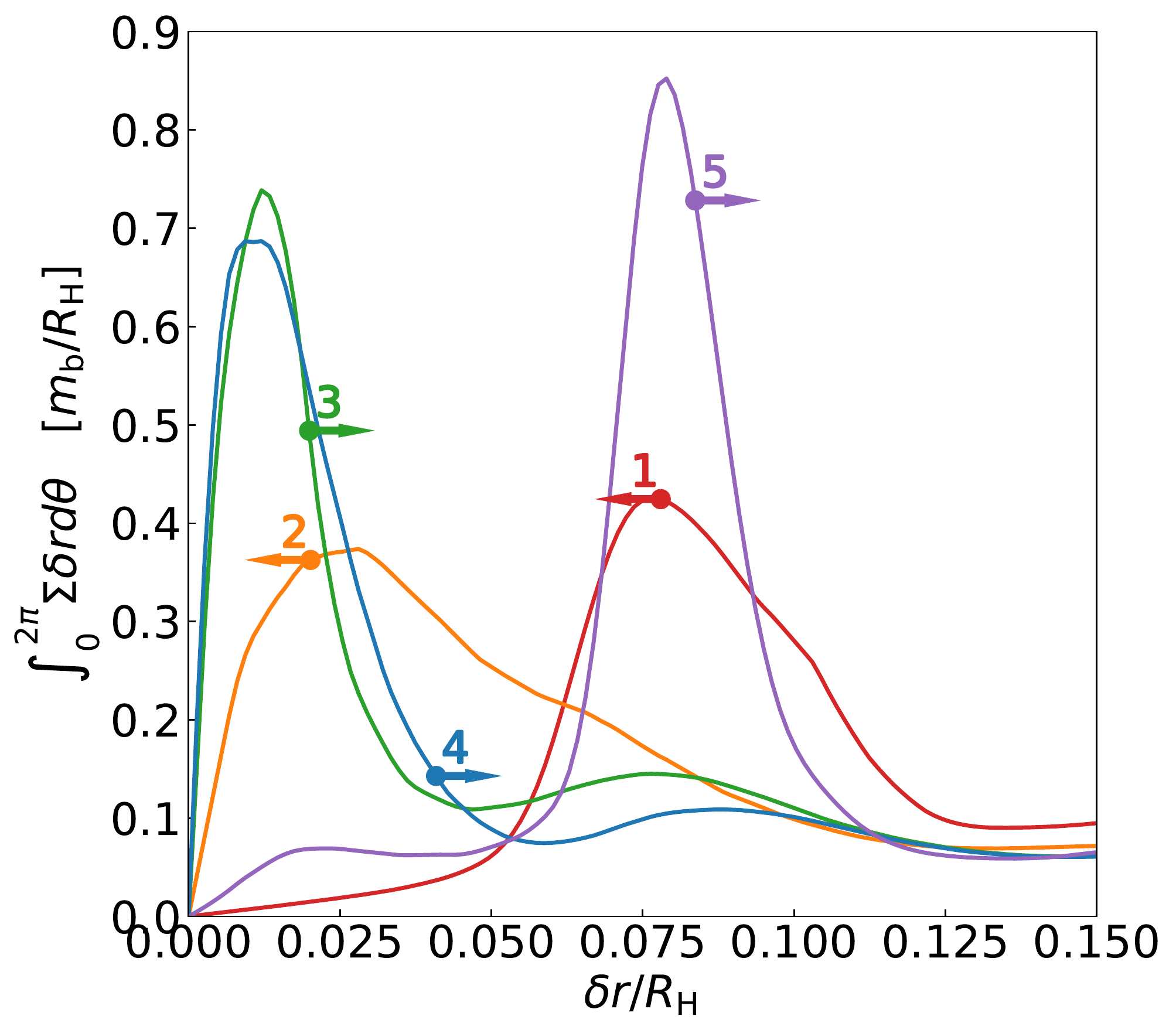}
    \epsscale{0.525}
    \plotone{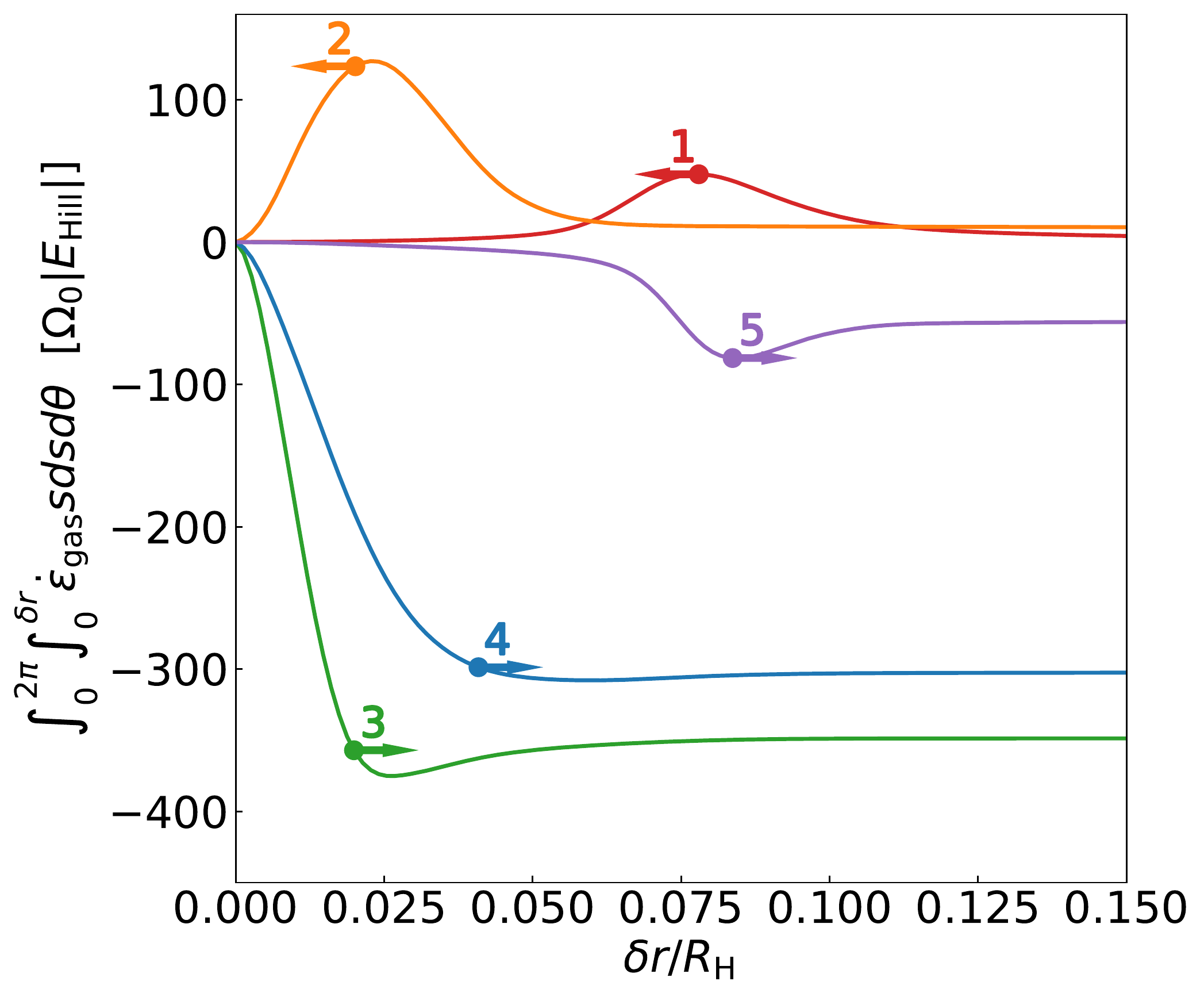}
    \caption{
    Time evolution of the binding energy for the simulation shown in Figure \ref{fig:2d-den-Edot} (top panel) as well as radial profiles of the gas linear mass density (bottom left) and cumulative power delivered to the binary (bottom right) at five different times (shown on the top panel). 
    Before the first close encounter (curves with left-pointing arrows), most of the gas tracks the BHs. After the close encounter (curves with right-pointing arrows), most of the mass remains near the interaction site ($\delta r\sim 0$) and removes a significant amount of energy from the BHs. This is in agreement with the 2D picture shown in Figure \ref{fig:2d-den-Edot}.  
    }
    \label{fig:three-analysis}
\end{figure*}

We consider a system with a supermassive black hole (SMBH) of mass $M$ and a Keplerian gas disk surrounding the SMBH. 
Inside this AGN disk, we embed two BHs of masses $m_1=10^{-5}M$ and $m_2=5\times10^{-6}M$.
Both $m_1$ and $m_2$ are initially placed on co-planar circular orbits around the SMBH, with orbital radii $a_1=R_0$ and $a_2 = a_1 + 2 R_{\rm H}$, where $R_0$ is a reference radius and  
\begin{eqnarray}
   R_{\rm H} &=& \frac{a_1+a_2}{2}\left(\frac{m_{\rm b}}{3M}\right)^{1/3} 
\end{eqnarray}
is the initial mutual Hill radius of the two BHs and where $m_{\rm b} = m_1 + m_2$. 
We use $\vec{r}_{1,2}$ to denote the position vectors of the BHs measured from the SMBH.
For our chosen parameters, $a_2 = 1.034 R_0$ and $R_{\rm H} = 1.017 R_0$. 

We model the disk in a two-dimensional cylindrical coordinate system ($R,\phi$) centered on the SMBH.
The disk is initialized with a constant surface density, $\Sigma(R,\phi)=\Sigma_0$. 
We approximate the gas temperature as isothermal with a constant sound speed $c_{\rm s}=0.01R_0\Omega_0$, where $\Omega_0=\sqrt{GM/R_0^3}$.
This setup gives a vertical disk scale height profile of
\be
H(R) = 0.01 R \sqrt{\frac{R}{R_0}}.
\ee
The disk is given a constant kinematic viscosity $\nu=10^{-6}R_0^2\Omega_0$, which corresponds to $\alpha=0.01$ at $R_0$ in the $\alpha$-disk model.

We simulate the evolution of our systems using a new version of the hydrodynamics code \texttt{LA-COMPASS} \citep{Li2005,Li2009} now coupled to the \texttt{REBOUND} $N$-body library \citep{Rein2012}. 
A detailed description of the numerical methods and the equations evolved is given in Appendix \ref{app:numerical}.

The focus of our simulations is on resolving the close encounters of the two BHs with the goal of determining under what conditions those close encounters end with a bound binary. 
To facilitate this, we limit the radial extent of the AGN disk to $[0.853,1.214]R_0$ so that we can have a high spatial resolution around each BH.
Additionally, we use a small softening length, $r_s \approx 0.0005 R_0$, so that each BH has a fully resolved circum-single disk (CSD) around itself. 
We do not consider gas accretion onto the BHs.
Our simulations are not designed to answer the question of how the BHs get into their initial close orbits with $a_2-a_1=2R_{\rm H}$. 
Answering that important question requires simulations that include a global model for the AGN disk and a model for how BHs are born or captured in the disk and how they migrate. 

The simulations include two stages: relaxation and release.
We perform relaxation in the first $500P_0$, where $P_0$ is the Keplerian orbital period around the SMBH at $R_0$.
During this time, the two BHs are fixed on their initial circular orbits to allow the gas to reach a quasi-steady-state distribution. 
The simulations run with a reduced, uniform resolution of $\sim50$ cells per $R_{\rm H}$.

After $500 P_0$, we double the resolution of the simulation and let the simulation adjust to the new resolution over the course of at least another $20 P_0$. 
Finally, we finish the relaxation and release the BHs when their azimuthal separation, $\varphi_2-\varphi_1$, is equal to a chosen value of $\Delta \varphi$.
Once released, the BHs move in response to the gravitational force of the other BH, SMBH, and the AGN disk. 
We run each simulation until a tightly bound binary has been formed or when the simulation reaches $t=540P_0$. 

In the next sections, we explore a range of values for $\Sigma_0$ and $\Delta \phi$ in order to build a suite of close encounter events with different impact parameters.

\section{Results}
\label{sec:result}

Figure~\ref{fig:r12_table} provides an overview of the outcomes of our main suite of simulations. 
In it, we show the time evolution of each BH's radial location in the disk.  
Rows correspond to a fixed $\Sigma_0$ while columns correspond to a fixed initial $\Delta \varphi$. 
We refer to AGN disks with $\Sigma_0 R_0^2= [10^{-5}, 5\times 10^{-5}, 10^{-4}] M$ as low, medium, and high mass, respectively.\footnote{For reference, for a single BH (mass $m_1$) moving in a disk, the characteristic eccentricity damping time is approximately $\tau_e=M^2h^4/(2\pi m_1\Sigma R_0^2)P_0$ \citep[e.g.,][]{Goldreich1980,Tanaka2004}. Our three disk densities correspond to
$\tau_e=16$, $3.2$ and $1.6P_0$, respectively. These $\tau_e$ values are much smaller than those considered in \cite{LJR2022ApJ}, which adopted a simple drag force formula to mimic the disk effect
with a drag time of $\gtrsim 10^5P_0$.}
Simulations are named according to their $\Delta \varphi$ and $\Sigma_0$ values, e.g., simulation \texttt{dp60high} has $\Delta \varphi = 60^\circ$ and $\Sigma_0$ corresponding to the high mass case. 

In most simulations, the two BHs experience their first close encounter at around $t=534P_0$.
As the figure shows, the outcome of that close encounter is dependent on both $\Delta \varphi$ and $\Sigma_0$. 
Lighter disks tend to not form binaries, and instead the BHs are kicked back to their initial orbital radii but now with a non-zero and small eccentricity. 
Binary formation does occur for the medium and high disk masses, but not at all $\Delta \varphi$. 
Among the high disk mass cases, binary formation occurs in \texttt{dp30high}, \texttt{dp60high}, and \texttt{dp90high}; among the medium disk mass cases, formation occurs in \texttt{dp60med}, \texttt{dp90med}, \texttt{dp120med}, and \texttt{dp150med}. 

Of the ``no-formation" simulations, there are a few oddities: \texttt{dp30med}, \texttt{dp120high}, and \texttt{dp180med}. The first, \texttt{dp30med} does not capture despite the trend seen at high masses of smaller $\Delta \varphi$ providing a better chance at capture. 
This ``non-monotonic'' dependence of the outcome on $\Delta \varphi$ could be related to the fractal structure seen in the gas-free simulations by \cite{Boekholt2022}.\footnote{\cite{Boekholt2022} showed that for $\Delta\varphi=180^\circ$, the smallest distance reached in close encounters between $m_1$ and $m_2$ depends sensitively on the initial $\Delta a=a_2-a_1$, and exhibits a fractal structure. We have carried out similar gas-free simulations for a fixed $\Delta a=2R_{\rm H}$, and found that the smallest distance also depends sensitively on the initial $\Delta\varphi$.}
The second, \texttt{dp120high} results in a BH with the largest orbital eccentricity of our sample ($e_{\rm bh} \sim 0.037$). 
This BH has $e_{\rm bh} \sim 3.5 H/a_{\rm bh}$, and so the eccentricity will likely be relatively long-lived since $\dot{e}_{\rm bh} \propto - e_{\rm bh}^{-2}$ in this regime \citep{Papaloizou2000MNRAS,Cresswell2007AandA,LiYP2019ApJ} as opposed to the faster, linear damping rate of $\dot{e}_{\rm bh}\propto - e_{\rm bh}$ \citep{Goldreich1980,Tanaka2004}.
The other oddity, \texttt{dp180med}, actually has an orbit-crossing, i.e., the BHs switch positions after the close encounter. 
The eccentricity of the inner BH also reaches a relatively large value of $e_{\rm bh} \sim 0.032$. 

All of the no-formation simulations could, in principle, have subsequent close encounters if they can reduce their mutual separations. 
Our simulations, however, are not designed to answer this question, as the limited radial extent of the AGN disk and the very small BH softening length are not ideal for determining the {\it global} disk migration of the BHs. 
We leave the answer to that question to future work, and instead turn our attention to the simulations which do form binary BHs.

\subsection{Formation of Binary BHs} 

For the seven formation cases depicted in Figure~\ref{fig:r12_table}, we provide a more detailed view of the binary formation process and the time-evolution of the BH separation and quantities that characterize the binary in Figure~\ref{fig:rael-vs-time}.
The orbit of a bound BBH is specified by its specific energy and angular momentum,
\be
E_{\rm b}  = \frac{1}{2} v_{\rm b}^2 - \frac{G m_{\rm b}}{r_{\rm b}} , \qquad \vec{\ell}_{\rm b} = \vec{r}_{\rm b} \times \vec{v}_{\rm b} ,
\ee
where the vectors $\vec{r}_{\rm b} = \vec{r}_1 - \vec{r}_2$ and $\vec{v}_{\rm b} = \vec{v}_1 - \vec{v}_2$ are the binary separation and binary velocity, respectively. 
From these quantities, we also calculate the orbital elements of each BBH as,
\be
a_{\rm b} = -\frac{G m_{\rm b}}{2 E_{\rm b}} , \qquad e_{\rm b}^2 = 1 - \frac{\ell_{\rm b}^2}{G m_{\rm b} a_{\rm b}}  .
\ee
An isolated bound binary requires $E_{\rm b} < 0$. 
However, binaries that are a part of a hierarchical triple have a stricter criterion of $E_{\rm b} < \chi E_{\rm Hill}$, where $E_{\rm Hill} = -G m_{\rm b}/(2 R_{\rm H})$ is the binding energy with $a_{\rm b} = R_{\rm H}$, and $\chi>1$ is an additional factor that depends on the binary eccentricity and mass ratio. 
Previous work on the stability of heirarchical triples find $\chi \sim 2 - 4$ for our parameters \citep{Eggleton1995ApJ,Mardling2001MNRAS,LiYP2021ApJ}. 

Close encounters are marked by a sharp dip in $r_{\rm b}$ where the BHs can be momentarily separated by one grid cell or less.
In all of the close encounters shown in Figure \ref{fig:rael-vs-time}, the binary always {\it loses} energy ($E_{\rm b}$ decreases) and $r_{\rm b}$ never exceeds $R_{\rm H}$ again. 
Outside of the close encounters, the binary energy is roughly conserved.  
From the second row of  Figure \ref{fig:rael-vs-time}, we see that while the energy lost per close encounter is relatively small, the binary experiences multiple close encounters within just one $P_0$ so that the cumulative loss of energy is great enough to turn a marginally bound binary (after the first close encounter) into a strongly bound binary. 

Examining the evolution of the binary eccentricity and angular momentum (third and fourth rows of Figure \ref{fig:rael-vs-time}), we find that in all cases, binaries capture with high eccentricity ($e_{\rm b} > 0.6$) and into retrograde orbits ($\vec{\ell_{\rm b}} \cdot \vec{\ell}_{\rm com} < 0$, where $\vec{\ell}_{\rm com}$ is the center-of-mass orbital angular momentum about the SMBH) . 
These results suggest that dynamically-assembled BBHs in AGN disks form in highly-eccentric, retrograde configurations with $a_{\rm b}/R_{\rm H} \sim 1/ {\rm ten} - 1/{\rm few}$. 
This is atypical of most of the studied embedded BBHs in the literature, as those are usually circular and prograde (e.g., \citealt{Samsing2020} and \citealt{LiYP2022ApJL}, but also see \citealt{LiYP2021ApJ} and \citealt{LiRX2022arxiv02} for a few retrograde examples).
But our binaries are circularizing on a timescale of about $P_0$ to $20P_0$, and will likely further contract on longer timescales \citep{LiYP2021ApJ}. 

\subsection{Anatomy of a Close Encounter}

In this section, we examine, in detail, the energy exchange between the BHs and the gas before, during, and after the initial close encounter for one exemplary simulation \texttt{dp30high}. 
We will show that the BHs lose a significant amount of energy to the gas that is unable to follow the BHs as they depart from the close encounter region. 
This gas imparts a significant drag force to the BHs, which we call a post-collision drag, helping them dissipate enough energy to bind. 

We start by showing in Figure \ref{fig:2d-den-Edot} snapshots of the gas density and power delivered to the binary at three times: at time of release, just before the close encounter, and after the close encounter. 
We define the power delivered to the binary as,
\be 
\dot{E}_{\rm gas} = \vec{v}_{\rm b} \cdot ( \vec{f}_{g,1} - \vec{f}_{g,2}) ,
\ee
where  $\vec{f}_{g,i}$ is the force from a given cell on BH $i$. 
From $\dot{E}_{\rm gas}$, we calculate the power per unit area as,
\be
\dot{\varepsilon}_{\rm gas} = \frac{d \dot{E}_{\rm gas}}{dA} ,
\ee
where $dA=s ds d\theta$ is the area element in a polar coordinate system $(s,\theta)$ centered on the geometric center of the binary. 

Well before the close encounter (left panels), each BH has collected gas into a well-defined CSD. 
Because of our small BH softening, the CSDs are resolved enough to see the spiral arms excited by the SMBH. 
We also see the trailing Lindblad spirals that extend away from the CSDs. 
As the BHs reach the close encounter (middle panels), their CSDs collide, causing a pile-up of material at the interaction site.  
The material at the center adds energy to the BHs ($\dot{\varepsilon}_{\rm gas} > 0$), whereas material exterior to the BHs tends to slow them down. 
As the BHs depart from the interaction site (right panels), most of the gas remains at the center, as it is unable to expand at a sufficient rate to keep up with the supersonic BHs. 
The material left behind now removes energy from the BHs ($\dot{\varepsilon}_{\rm gas} < 0$), and since most of the material is in this region, the total $\dot{E}_{\rm gas}$ at this stage is negative. 

We further demonstrate this last point in Figure \ref{fig:three-analysis}, where we plot the radial ($\delta r$) profiles of the linear mass density ($\int_0^{2 \pi} \Sigma\delta r d\theta$), the cumulative energy deposition rate ($\int_0^{2 \pi}\int_0^{\delta r} \dot{\varepsilon}_{\rm gas} s ds d\theta$), and the binary separation at five times before and after the close encounter. 

Before the close encounter at events 1 and 2, most of the mass tracks the BHs as they approach each other.
During this time, gas interior to the BHs increases $\dot{E}_{\rm gas}$ while gas exterior to the BHs removes energy, but in total there is very little change to the BHs binding energy. 
The situation changes after the close encounter (events 3 and 4). 
Now, most of the mass is near the center -- far inside of the BHs. In fact, at event 4, almost all of the mass is inside of $0.025 R_{\rm H}$. 
Furthermore, this gas stuck at the center removes a large amount of energy from the BHs -- far more than what was gained before the close encounter. 
To reiterate, because this material cannot keep up with the BHs, the gravitational interaction acts as a drag force trying to slow the BHs departure down. 
By the time of event 5, most of the mass has caught up to the BHs, and there is only a small amount of drag remaining. 

We see that from the top panel of Figure \ref{fig:three-analysis}, this process is more-or-less repeated at subsequent close encounter events. 
The strength of each successive close encounter is progressively diminished, but altogether the amount of energy dissipated is enough to take $E_{\rm b}$ from $-|E_{\rm Hill}|$ to $\lesssim -9 |E_{\rm Hill}|$. 

Although our analysis here is based on one example, this post-collision drag can be expected to arise in other BH close encounters. 
Due to the highly non-linear and non-uniform gas density distribution, this drag force is drastically different from standard gas dynamical friction \citep[e.g.,][]{Papaloizou2000MNRAS}.

\section{Summary and Discussion}
\label{sec:summary}
\subsection{Summary}

Our main results are shown in Figures \ref{fig:r12_table} and \ref{fig:rael-vs-time}. 
We have shown that dynamical formation of binary BHs in AGN disks is possible for sufficiently high disk density and occurs in a three-stage process: 
\begin{enumerate}
 \item Two BHs on isolated orbits around a SMBH come within $2 R_{\rm H}$ of each other and have a close encounter where their separation momentarily goes below a small fraction of $R_{\rm H}$. The binding energy of the two BHs at this point needs to be somewhat smaller than $E_{\rm Hill}=-Gm_{\rm b}/(2R_{\rm H})$.  
 \item The circum-single disks around each BH collide and disrupt as the BHs scatter. 
 \item The remnant material left behind  reduces the relative departure speed of the BHs enough to capture them into a bound binary. 
\end{enumerate}

The binaries formed by this mechanism tend to be {\it highly eccentric} and {\it retrograde} with respect to their center-of-mass motion around the SMBH. 
Our findings suggest that formation (via one scattering event) in gas-rich environments is much more likely than in gas-free or gas-poor environments \citep{LJR2022ApJ}. 

The post-collision drag mechanism relies on the gas that was in the CSDs being unable to follow the BHs as they scatter off of each other. 
Because our simulations are isothermal, we do not account for the gas heating during close encounters.
Therefore, the gas at the close encounter site is both denser and colder than it would be in an adiabatic simulation.
Both of these effects mean that isothermal simulations may overpredict the amount of drag on the BHs. 
Future work using more realistic thermal physics should investigate this effect.

\subsection{A close encounter merger criterion for $N$-body simulations}
\label{sec:conditions}
\begin{figure}[t]
    \includegraphics[width=.47\textwidth]{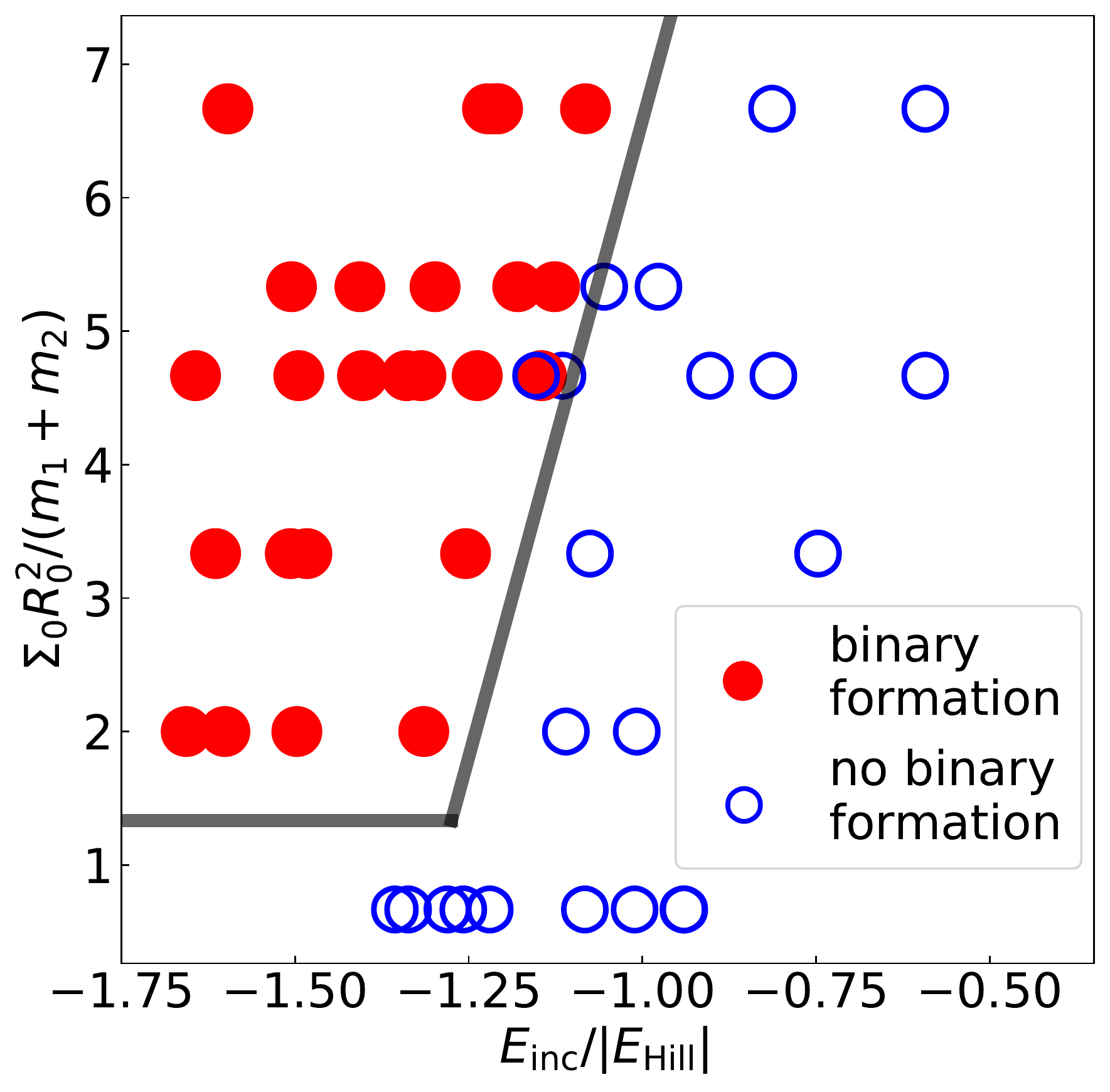}
    \caption{
    Scatter plot of close encounter outcomes for our expanded set of simulations. 
    The y-axis encodes the mass scale of the AGN disk, while the x-axis shows the value of $E_{\rm b}$ just before the first close encounter (evaluated when $r_{\rm b}=0.3 R_{\rm H}$) in units of $|E_{\rm Hill}|=Gm_{\rm b}/(2R_{\rm H})$.  
    We color each point by whether or not the BHs catch into a bound system. 
    In this space there is a clear dividing line between formation and no-formation that we mark with a grey line and provide a numerical fit for in Equation \eqref{eq:condition}. 
    }
    \label{fig:parameter-space}
\end{figure}

Many previous studies about the long-term evolution of BHs in AGN disks do not capture the effects of the gas disk with high fidelity.
Instead, these are typically pure $N$-body calculations that may have idealized approximate gas effects such as migration or drag. 
As we have shown, though, the treatment of close encounters is critical to binary formation. 
We have run an extended suite of simulations at different $\Sigma_0$ and $\Delta \varphi$ values with the goal of providing a new merger criterion for close encounters. 

In Figure \ref{fig:parameter-space}, we show the outcome of each simulation (formation or no-formation) for two ``input'' parameters:  $\Sigma_0 R_0^2/m_{\rm b}$ and the two-body energy just before the close encounter $E_{\rm inc} \equiv E_{\rm b}(r_{\rm b} = 0.3 R_{\rm H})$. 
In this parameter space, there is a clear separation between formation and no-formation. 
Formation occurs when the disk-to-binary mass is above a critical value that is a function of the incoming close encounter energy. 
From Figure \ref{fig:parameter-space}, we empirically determine the binary formation criterion to be, 
\be
\label{eq:condition}
\left(\frac{\Sigma_0 R_0^2}{m_1+ m_2} \right) > \mu_{\rm crit} = {\rm max}(1.3, 19.1 E_{\rm inc}/|E_{\rm Hill}| + 25.6). \nonumber \\ 
\ee
This can then be used as a ``merger'' criterion in place of a purely separation-dependent criterion (e.g., $r_{\rm b} < R_{\rm H}$) or a simple binding energy criterion (e.g., $E_{\rm b} < 0$), as these previous criteria tend to over-predict the formation rate of binaries.

We note that Figure~\ref{fig:parameter-space} and Equation~\eqref{eq:condition} apply to our simulations with $m_{\rm b}/M=1.5\times10^{-5}$ and $H/R\approx0.01$.
We also note that this criterion might change at much higher $\Sigma_0 R_0^2/m_b$ and $E_{\rm inc}/|E_{\rm Hill}|$.
More studies will be needed to explore how the result changes for different parameters.

\subsection{Future directions} 
There are several fruitful directions of research that can expand upon and improve our results. Here, we just provide a cursory summary. 

3D effects, including the vertical structure of the AGN disk and CSDs, as well as mutual inclinations between the BHs will alter the gravitational and hydrodynamical details of the close encounters.  
Moreover, non-isothermal equations of state, radiative processes, and self-gravity of the gas can all affect the gas density in the CSDs.
Gas accretion onto the BHs and magnetohydrodynamics of the AGN disks can also play a role.
Future simulations may incorporate these effects. 

Follow-up studies should explore the long-term evolution of the newly-formed binaries (e.g., timescales for circularization and merger) and the BHs in the no-formation cases (e.g., probability of undergoing more close encounters in the future).

Future simulations can also examine this formation mechanism in a larger parameter space (e.g., different AGN disk mass and temperature, BH masses and orbital elements) and further generalize the `merger' criterion (Equation~\ref{eq:condition}).
Then, long-term $N$-body simulations with many BHs in a disk can apply the `merger' criterion to resolve close encounters. 
The result will provide insight to the rate and properties of the gas-captured BBHs in AGN disks.

\acknowledgments
We gratefully acknowledge the support by LANL/LDRD under project number 20220087DR.
This work is also supported in part by NSF grant AST-2107796 and the NASA grant 80NSSC19K0444.
Jiaru Li is also supported by the Center for Space and Earth Science at Los Alamos National Laboratory.
This research used resources provided by the Los Alamos National Laboratory Institutional Computing Program, which is supported by the U.S. Department of Energy National Nuclear Security Administration under Contract No. 89233218CNA000001.
This work is approved for unlimited release under LA-UR-22-29010.

\software{
\texttt{LA-COMPASS} \citep{Li2005,Li2009},
\texttt{Rebound} \citep{Rein2012,Rein2015},
\texttt{Numpy} \citep{Walt2011},
\texttt{Scipy} \citep{Virtanen2020},
\texttt{Matplotlib} \citep{Hunter2007}.
}

\appendix
\section{Numerical Method}
\label{app:numerical}
We briefly describe the new version of the hydrodynamics code \texttt{LA-COMPASS} \citep{Li2005,Li2009} here.
\texttt{LA-COMPASS} is a finite-volume, Godunov-type code that, in its most general form, solves the MHD equations coupled to a self-gravity solver, a multi-species dust fluid, and a population of planetary-sized bodies.
For this work we do not follow the evolution of dust nor do we include magnetic fields or self-gravity. 
For the gas in the AGN disk, \texttt{LA-COMPASS} evolves the 2D gas surface density ($\Sigma$) and velocity ($\vec{v}$) according to,  
\be
\frac{\partial \Sigma}{\partial t} + \nabla \cdot (\Sigma \vec{v}) 
& = & 
0 \\
\frac{\partial (\Sigma \vec{v})}{\partial t} + \nabla \cdot (\Sigma\vec{v}\vec{v}) 
& = &
-\Sigma\nabla\Phi - \nabla P +  f_{\nu},
\ee
where $P$ is the gas pressure, $\Phi$ is the total external gravitational potential, and $f_{\nu}$ is the viscous force.
We assume that the gas is isothermal so that $P=c_s^2 \Sigma$. 
The gravitational potential contains the SMBH potential and the two BH potentials,
\be
\Phi(\vec{R}) = -\frac{GM}{R} - \sum_{i=1,2}\frac{Gm_i}{r_{s,i}}\times S\left(\frac{|\vec{r}_i-\vec{R}|}{r_{s,i}}\right),
\ee
where $r_{s,1} = 0.0005 R_0$ and $r_{s,2}= 0.000526 R_0$ are the softening lengths and
\be
S\left(x\right) =
\begin{cases}
x^3 - 2x^2 +2 & \text{for } x\leq1\\
1/x & \text{for } x > 1 
\end{cases}
\ee
is the spline-softening function \citep{Kley2009AandA}.

The BHs and SMBH evolve under the action of their mutual gravity plus the total gravitational force from the disk. 
In previous versions, \texttt{LA-COMPASS} solved the $N$-body system with a high order Runge Kutta scheme. 
In this paper, we have replaced this integrator with the full, $N$-body code  \texttt{REBOUND} \citep{Rein2012}.
\texttt{REBOUND} offers a variety of integrators designed for planetary motion. 
We have chosen the 15th-order \texttt{IAS15} integrator for this work \citep{Rein2015}. 
We assume the disk force remains constant in one hydro step. \texttt{REBOUND} may use sub-cycling and take more time steps to satisfy the error tolerance in one hydro step, which we set to $10^{-9}$. Although, we have found that the \texttt{REBOUND} time step does not go below the hydro time step -- even when the BHs have a close encounter.


\end{document}